\documentclass[aps,preprint]{revtex4}

\usepackage{epsfig}

\def\bea{\begin{eqnarray}}
\def\eea{\end{eqnarray}}
\def\bec{\begin{center}}
\def\ec{\end{center}}

\def\beq{\begin{equation}}
\def\eeq{\end{equation}}

\begin{document}
\draft
\tighten
\preprint{KAIST-TH 2005/12}
\preprint{KUNS-1982}
\preprint{KYUSHU-HET-85}
\title{\large \bf Little SUSY hierarchy in mixed modulus-anomaly mediation}
\author{
Kiwoon Choi\footnote{kchoi@hep.kaist.ac.kr}$^1$,
Kwang Sik Jeong\footnote{ksjeong@hep.kaist.ac.kr}$^1$,
Tatsuo Kobayashi\footnote{kobayash@gauge.scphys.kyoto-u.ac.jp}$^2$ and
Ken-ichi Okumura\footnote{okumura@higgs.phys.kyushu-u.ac.jp}$^3$
}
\address{
$^1$Department of Physics, Korea Advanced Institute of Science
and Technology, Daejeon 305-701, Korea \\
$^2$Department of Physics, Kyoto University,
Kyoto 606-8502, Japan \\
$^3$Department of Physics, Kyushu University, Fukuoka 812-8581,
Japan}
\begin{abstract}
Motivated by the KKLT string compactification involving a supersymmetry-breaking
uplifting potential, we examine 4D effective supergravity
with a generic form of uplifting potential, focusing
on the possibility that the resulting mixed modulus-anomaly mediated soft terms
realize the little hierarchy between the Higgs boson
masses $m_H$  and the sparticle masses $m_{SUSY}$.
It is noted that for some type of uplifting potential,
the anomaly-mediated contribution
to $m_H^2$ at $M_{GUT}$
can cancel the subsequent renormalization group
evolution of $m_H^2$ down to TeV scale,
thereby the model
can naturally realize the little hierarchy $m_H^2\sim m_{SUSY}^2/8\pi^2$
which is desirable  for the lightest Higgs boson mass
to satisfy the experimental bound.
In such models, the other Higgs mass parameters $\mu$ and $B$
can have the desirable size $\mu \sim B \sim m_H$
without severe fine-tuning of parameters, although
the gravitino is much heavier than the Higgs boson.
Those models for the little hierarchy avoid naturally the
dangerous SUSY flavor and CP violations, and
predict nearly degenerate low energy gaugino masses, pure Higgsino LSP,
and also a specific relation between
the stop and gaugino masses.
\end{abstract}
\maketitle


Low-energy supersymmetry (SUSY) is one of primary candidates
for physics
beyond the standard model (SM) above the weak scale \cite{Nilles:1983ge}.
One of the most important motivations for supersymmetric extension
of the SM is to solve the hierarchy problem
between the weak scale and GUT/Planck scale,
The minimal supersymmetric extension of the
standard model is important from the viewpoint of its
minimality as well as the realization of gauge coupling
unification at $M_{GUT} \sim 2 \times 10^{16}$ GeV.
However, the minimal supersymmetric standard model (MSSM)
seems to face a fine-tuning problem,
the so-called little SUSY hierarchy problem \cite{Barbieri:1987fn}.

The little SUSY hierarchy problem is caused by the
combination of the following aspects of the MSSM.
First of all, the experimental
bound of the CP-even Higgs boson mass $m_{h^0}$ requires
a rather large stop mass, $m_{\tilde t} \geq 500$ GeV,
to enhance the radiative correction to $m_{h^0}^2$ due to
the top-stop mass splitting \cite{Haber:1990aw}.
On the other hand, the soft SUSY breaking scalar mass of
the up-sector Higgs field $m_{H_u}$ has a renormalization group (RG) evolution due to
$m_{\tilde{t}}$:
\begin{equation}
\Delta m_{H_u}^2 \sim - \frac{3}{4\pi^2}y_t^2
m_{\tilde t}^2 \ln \frac{\Lambda}{m_{\tilde t}},
\label{log-h-mass}
\end{equation}
where $y_t$ is the top Yukawa coupling and $\Lambda$ is the cut-off scale.
This RG evolution effect indicates that
$|m_{H_u}^2|$ at TeV scale is generically  close to
$m_{\tilde t}^2$ for $\Lambda \sim 10^{16}$ GeV.
Finally the minimization condition of the
Higgs potential in the MSSM leads to
\begin{equation}
\frac{M_Z^2}{2} \simeq - \mu^2 - m^2_{H_u},
\label{MZ-mu-mHud}
\end{equation}
for a moderate and large value of $\tan \beta$.
(This approximation is valid even for
$\tan \beta \sim 3$ when $|m^2_{H_d}| \sim |m^2_{H_u}|$.)
Then for
$m_{H_u}^2 \sim m_{\tilde{t}}^2\geq {\cal O}(500^2)$ GeV$^2$,
one should fine-tune $\mu^2$
in order to derive the weak scale,
and the required degree of fine-tuning is
${\cal O}(1\%)$ or more severe.
That is the little hierarchy problem.

Recently, several types of scenarios
extending the MSSM
\cite{casas}-- \cite{Delgado:2005fq}
have been proposed
to solve the little hierarchy problem.
From a simple bottom-up viewpoint,
a favored pattern of mass parameters would be
\begin{equation}
m_{\tilde t}^2 \,\gg\,
|m_{H}^2| \,\sim\, \mu^2 \,\sim\, |\mu B|
\,=\,
{\cal O}(100^2){\rm~GeV}^2,
\label{mass-parameter}
\end{equation}
at low energy scale.
Then the key-point to solve the little hierarchy problem
is to achieve (\ref{mass-parameter}) with canceling the
large radiative correction (\ref{log-h-mass}).
One scenario based on superconformal dynamics has been
proposed \cite{Kobayashi:2004pu},
in which the superconformal dynamics cancels
the large logarithmically divergent corrections,
while leaving only small finite corrections.
Here we propose another scenario based on the particular feature
of the mixed modulus-anomaly mediated SUSY breaking scenario which might be
realized in KKLT-type string compactification with SUSY-breaking
anti-brane \cite{Kachru:2003aw,choi1}.


Superstring theory is a promising candidate for unified theory
including gravity.
However, compactified string theory
in general includes moduli fields
which have a flat potential perturbatively.
How to stabilize those moduli
has been one of the most outstanding issues in string phenomenology.
Recently, KKLT has proposed a new scenario
to stabilize moduli and break SUSY in type IIB string compactification
\cite{Kachru:2003aw}.
All complex structure moduli and the IIB dilaton are stabilized
by the effects of 3-form fluxes.
On top of that, the remaining K\"ahler moduli are stabilized
through non-perturbative dynamics at SUSY AdS vacuum, and
finally the vacuum is lifted to a dS (or Minkowski) vacuum by
uplifting potential induced by
anti-$D3$ brane.
Soft SUSY breaking terms in such scenario has been
studied in Ref.~\cite{choi1}, and it has been shown that
a quite new pattern of soft terms arises.
In KKLT scenario, K\"ahler moduli $F$-terms are of the order of
$m_{3/2}/4\pi^2$, thereby the moduli $F$-terms \cite{modulimediation}
and anomaly mediation \cite{Randall:1998uk} contribute comparably
to the resultant soft masses $m_{SUSY}\sim m_{3/2}/4\pi^2$.
Under a reasonable condition, anomaly mediated contributions
 at $M_{GUT}$
cancel the subsequent RG evolution of soft parameters between
$M_{GUT}$
and a {\it mirage messenger scale} $\Lambda_{m}\sim (m_{3/2}/M_{Pl})^{\alpha/2}M_{GUT}$
where $\alpha$ is a parameter of order unity which is determined
by the moduli-dependence of uplifting potential
\cite{Choi:2005uz}.(See also Refs.~\cite{Endo:2005uy,Falkowski:2005ck}
for other phenomenological aspects.)
In the original KKLT model \cite{Kachru:2003aw},
the uplifting potential from anti-$D3$ brane  gives $\alpha=1$ \cite{choi1}.
However different forms of uplifting potential
might be possible in other type of string compactifications, yielding different value of  $\alpha$.
As we will discuss in this paper,
the cancellation of RG evolution
in the mixed modulus-anomaly mediation can have an interesting implication
for the little hierarchy problem, particularly
for certain form of uplifting potential yielding $\alpha=2$.

In this paper, we examine the possibility of realizing the little
hierarchy $m_H^2\sim m_{SUSY}^2/8\pi^2$
in the framework of 4D effective supergravity (SUGRA)
with a general form of uplifting potential.
We discuss also the Higgs mass parameters
$\mu$ and $B$ in such effective SUGRA.
Generically for the MSSM embedded in KKLT-motivated SUGRA,
the resulting $B$ is of the order of  $m_{3/2}\sim 4\pi^2 m_{SUSY}$,
thus too large to allow the correct electroweak symmetry breaking.
However the models with $\alpha=2$ allow  $B$
to be comparable to $m_H$ without severe fine-tuning of parameters,
as well as being able to lead to the natural cancellation of RG mixing between $m_H^2$ and $m_{SUSY}^2$.
Moreover, those models for the little hierarchy avoid naturally the dangerous SUSY flavor and CP violations,
and provide highly distinctive low energy predictions.



Although it appears to break SUSY explicitly, uplifting potential
induced by anti-brane can be accommodated in 4D SUGRA in a consistent manner
through a spurion operator (or more generally
through a non-linear Goldstino superfield) in $N=1$ superspace \cite{choi1}.
The effective action of such 4D SUGRA
can be written as
\begin{eqnarray}
\label{N=1}
S_{\rm eff}&=&\int d^4x \sqrt{g^C} \,\left[\,
\int d^4\theta \,
CC^*\Big(\,-3\exp(-K/3)\,\Big) -C^2C^{2*}\theta^2\bar{\theta}^2 {\cal P}_{\rm lift}
\right.\nonumber \\
&&+\,\left.\left\{
\int d^2\theta
\left(\frac{1}{4}f_a W^{a\alpha}W^a_\alpha
+C^3W\right)
+{\rm h.c.}\right\}\,\right],
\end{eqnarray}
where $K$, $W$, and $f_a$ denote the K\"ahler potential,
superpotential and gauge kinetic functions of the
standard $N=1$ 4D SUGRA, while
${\cal P}_{\rm lift}$ stands for the spurion operator
providing the uplifting potential.
Here we are using the superconformal formulation of 4D SUGRA
with chiral compensator superfield $C$,
and $g^C_{\mu\nu}$ is the 4D metric
in superconformal frame which is related to the Einstein frame metric
$g^E_{\mu\nu}$ as
$g^C_{\mu\nu}=(CC^*)^{-1}e^{K/3}g^E_{\mu\nu}$.
Note that one needs an off-shell formulation of $N=1$ SUGRA
in order to describe the coupling between the standard $N=1$ SUGRA sector
and the SUSY-breaking
anti-brane which is presumed to generate the uplifting
potential (or more generally a sector in which $N=1$ SUSY is non-linearly realized).
For simplicity, we choose the superconformal gauge in which
both the fermionic component of $C$ and the scalar auxiliary component
of SUGRA multiplet are vanishing, and then ignore the dependence
of SUGRA multiplets other than the metric dependence.
There still remains a residual super Weyl invariance under
\bea
C\rightarrow e^{-2\tau}C,
\quad
g^C_{\mu\nu}\rightarrow e^{2(\tau+\tau^*)}g^C_{\mu\nu},
\quad
\theta^\alpha \rightarrow e^{-\tau+2\tau^*}\theta^\alpha,
\eea
where $\tau$ is a complex constant, and the spurion operator
should be invariant under this super Weyl transformation
to keep the consistency of superconformal formulation.

Let $T$ denote a modulus superfield whose VEV determines
the unified gauge coupling at $M_{GUT}$, i.e.
\bea
f_a = T,
\eea
for the gauge kinetic functions of the SM gauge fields,
and assume that $T$ is the only light modulus which participates significantly
in SUSY-breaking.
We further assume that the theory possesses an approximate nonlinear PQ symmetry
\bea
\label{u1t}
U(1)_T : \quad T\rightarrow T + i\beta_T
\quad (\beta_T = \mbox{real constant}),
\eea
which is broken by nonperturbative dynamics stabilizing $T$.
Then the K\"ahler potential, superpotential and the spurion operator
can be written as
\bea
K&=&K_0(T+T^*)+Z_i(T+T^*)\Phi_i^*\Phi_i,
\nonumber \\
W&=& W_0(T)+\frac{1}{6}\lambda_{ijk}\Phi_i\Phi_j\Phi_k,
\nonumber \\
{\cal P}_{\rm lift} &=& {\cal P}_{\rm lift}(T+T^*),
\eea
where  $\lambda_{ijk}$ are $T$-independent constants and the $T$-dependence
of $W_0$ arises from non-perturbative dynamics.
In the original KKLT model realized in type IIB string theory,
$T$ corresponds to a K\"ahler modulus which represents the volume
of a 4-cycle wrapped by $D7$ branes containing the SM gauge fields.
Throughout this paper, we will focus on the KKLT-form of the modulus superpotential:
\bea
\label{kklt_super}
W_0 &=& w_0-Ae^{-aT},
\eea
where $w_0$ is a (flux-induced) constant and
$Ae^{-aT}$ is generated by non-perturbative dynamics such as
stringy instanton and/or field theoretic gaugino condensation.
We assume that $w_0={\cal O}(m_{3/2}M_{st}^2)$ is small enough to give low energy SUSY, while
$A={\cal O}(M_{st}^3)$ or ${\cal O}(M_{GUT}^3)$ for the
string or GUT scale which is rather close to $M_{Pl}$.
Note that using the $U(1)_R$ transformation:
$C\rightarrow e^{i\beta_R}C,
\quad W\rightarrow e^{-3i\beta_R} W
$
together with the non-linear PQ transformation
(\ref{u1t}), one can always make $w_0$ and $A$ to be real parameters.

It is  obvious that the spurion operator in (\ref{N=1})
does not affect the standard on-shell relations for
the SUSY breaking auxiliary components (in the Einstein frame):
\begin{eqnarray}
\label{approx-F}
\frac{F^C}{C_0}&= &\frac{1}{3}\partial_TK_0F^T+m_{3/2}^*,
\nonumber \\
F^T&= &-e^{K_0/2}\left(\partial_T\partial_{T^*}K_0\right)^{-1}\left(D_T
W_0\right)^*,
\eea
where  $C=C_0+\theta^2F^C$, $m_{3/2}=  e^{K_0/2}W_0$ and $D_TW_0=\partial_TW_0+W_0\partial_TK_0$.
On the other hand, the modulus potential  is modified to include
the uplifting term:
\bea
\label{approx-potential}
V_0&=&V_F+V_{\rm lift}
\nonumber \\
&=&e^{K_0}\left(\left(\partial_T\partial_{T^*}K_0\right)^{-1}D_TW_0(D_TW_0)^*-3|W_0|^2\right)+e^{2K_0/3}{\cal P}_{\rm lift}(T,T^*),
\eea
where $V_F$ is the standard $F$-term potential in $N=1$ SUGRA.
It is then straightforward to compute
$F^T$ and $F^C$  by
minimizing the above modulus potential under the fine tuning
for $\langle V_0\rangle =0$.
We then find the following relations between the VEVs:
\bea
\label{vev}
aT &=&\Big[
1+{\cal O}(\epsilon)\Big]\ln ({M_{Pl}}/{m_{3/2}}),
\nonumber \\
\frac{m_{3/2}}{M_0} &=&
\frac{(T+T^*)\partial_TK_0}{3\partial_T \ln(V_{\rm lift})}
\left[ a-\left(
\partial_TK_0
+\frac{\partial^2_TK_0}{\partial_TK_0}
-\frac{3\partial^2_TK_0\partial_T\ln(V_{\rm lift})}{
(\partial_T K_0)^2}\right)\right]
\Big[ 1+{\cal O}(\epsilon^2)\Big]
\nonumber \\
&=&\frac{2}{3}\frac{\partial_TK_0}{\partial_T\ln(V_{\rm lift})}
\Big[1+{\cal O}(\epsilon)\Big]\ln ({M_{Pl}}/{m_{3/2}}),
\eea
where $\epsilon=1/\ln(A/w_0)\sim 1/\ln(M_{Pl}/m_{3/2})$ is used as a small
expansion parameter and
\bea
M_0 \equiv  \frac{F^T}{(T+T^*)}.
\eea
Note that $\epsilon \sim 1/4\pi^2$ for $m_{3/2}$ in TeV range.
Generically
$\partial_T K_0/\partial_T \ln (V_{\rm lift})$
is of order unity, and thus the above result indicates
$m_{3/2}/M_0={\cal O}(\ln(M_{Pl}/m_{3/2}))={\cal O}(1/4\pi^2)$
{\it independently} of the detailed forms of $K_0$ and $V_{\rm lift}$.
As a result, if the modulus which determines
the unified gauge coupling is stabilized by the
KKLT-type superpotential (\ref{kklt_super}) and the vacuum is
lifted to a dS (Minkowski) state by a SUSY-breaking
spurion operator,
it is a {\it generic} consequence of the model that
the anomaly-mediated soft masses
 $\sim m_{3/2}/4\pi^2$ \cite{Randall:1998uk} are comparable  to the modulus-mediated
soft masses $\sim M_0$.


The soft terms in the above type of effective SUGRA have been
studied in \cite{choi1}.
For the soft terms of canonically normalized
visible fields:
\begin{eqnarray}
{\cal L}_{soft}&=&-\frac{1}{2}M_a\lambda^a\lambda^a-m_i^2|\phi_i|^2
-\frac{1}{6}A_{ijk}y_{ijk}\phi_i\phi_j\phi_k+{\rm h.c.},
\end{eqnarray}
where $\lambda^a$ are gauginos, $\phi_i$ are sfermions,
and $y_{ijk}=\lambda_{ijk}/\sqrt{e^{-K_0}Z_iZ_jZ_k}$ denote the canonically normalized
Yukawa couplings,
one finds the following mixed modulus-anomaly mediated soft parameters
at energy scale just below the unification scale \cite{choi1}:
\begin{eqnarray}
\label{soft1}
M_a
&=& M_0+\frac{b_a}{8\pi^2}g^2_{GUT}m_{3/2},
\nonumber \\
A_{ijk}
&=&\tilde{A}_{ijk}
-\frac{1}{16\pi^2}(\gamma_i+\gamma_j+\gamma_k)m_{3/2},
\nonumber \\
m_i^2
&=&\tilde{m}_i^2
-\frac{1}{32\pi^2}\frac{d\gamma_i}{d\ln\mu}m_{3/2}^2
\nonumber \\
&+&\frac{1}{4\pi^2}\left\{
\sum_{jk}\tilde{A}_{ijk}\left|\frac{y_{ijk}}{2}\right|^2-
C_2(\Phi_i)M_0\right\}m_{3/2},
\end{eqnarray}
where $\tilde{A}_{ijk}$ and $\tilde{m}_i^2$ are the
pure modulus-mediated trilinear $A$-parameters and soft scalar masses at $M_{GUT}$:
\bea
\label{puremodulus}
\tilde{A}_{ijk}&=&
(a_i+a_j+a_k)M_0,
\nonumber \\
\tilde{m}_i^2&=&\frac{2}{3}(V_F+V_{\rm lift})+c_i|M_0|^2,
\eea
for
\bea
a_i &=&(T+T^*)\partial_T\ln (e^{-K_0/3}Z_i),
\nonumber \\
c_i &=& -(T+T^*)^2\partial_T\partial_{T^*}\ln (e^{-K_0/3}Z_i),
\eea
and $C_2(\Phi_i){\bf 1}=\sum_a g_a^2T_a^2(\Phi_i)$ for the gauge generator $T_a(\Phi_i)$.
Here $b_a$ and $\gamma_i$ are the one-loop
beta function coefficients and the anomalous dimension of $Q_i$, respectively,
defined by
$\frac{dg_a}{d\ln \mu}=\frac{b_a}{8\pi^2} g_a^3$
and $\frac{d\ln Z_i}{d\ln \mu}=\frac{1}{8\pi^2}\gamma_i$.

Note that in the presence of the uplifting potential,
the modulus-mediated soft scalar mass is given by
\bea
\tilde{m}_i^2&=&\frac{2}{3}\left(V_F+V_{\rm lift}\right)+c_i|M_0|^2
\nonumber \\
&=& \Big[V_F+ m_{3/2}^2-F^TF^{T*}\partial_T\partial_{T^*}\ln (Z_i)\Big]
+\frac{2}{3}V_{\rm lift},
\eea
where the terms in the bracket of the second line
correspond to the modulus-mediated soft scalar mass in the standard $N=1$ SUGRA without
uplifting potential \cite{modulimediation},
and the last term is the contribution from uplifting spurion
which can be  determined only in the superspace
(off-shell) description of the uplifting potential such as in (\ref{N=1}).
Inclusion of this additional contribution is crucial for the correct calculation of soft scalar masses.
If not included, $\tilde{m}_i$ appears to be
of the order of $m_{3/2}\sim 4\pi^2M_0$, while the correct value
of $\tilde{m}_i$ is of the order of $M_0$ under the condition of
vanishing vacuum energy density: $\langle V_F+V_{\rm lift}\rangle=0$.
In general, any  source of
the vacuum energy density can affect soft scalar mass also, and
its contribution should be taken into account
for the correct evaluation of soft scalar mass \cite{ckn}.

In fact, at the leading approximation ignoring higher order (stringy) threshold
corrections, it is expected that $K_0,Z_i$ and ${\cal P}_{\rm lift}$ take a form:
\bea
\label{model}
K_0&=&-n_0\ln(T+T^*),
\nonumber \\
Z_i &=& \frac{1}{(T+T^*)^{n_i}},
\nonumber \\
{\cal P}_{\rm lift}&=& D(T+T^*)^{n_P},
\eea
where $n_0$, $n_i$ and $n_P$ are appropriate  {\it rational} numbers,
and $D$ is a constant to be adjusted for
$\langle V_F+V_{\rm lift}\rangle =0$.
For this form of $K_0,Z_i$ and ${\cal P}_{\rm lift}$,
we find
\bea
\label{rational}
\alpha &\equiv & \frac{m_{3/2}}{M_0\ln(M_{Pl}/m_{3/2})}
=\frac{2n_0}{2n_0-3n_P},
\nonumber \\
a_i &=& c_i = \frac{n_0}{3}-n_i,
\eea
up to small corrections
of ${\cal O}(1/8\pi^2)$.
In the original KKLT model, $n_0=3$ and
 the uplifting spurion originates from  anti-$D3$
brane for which $n_P=0$, and thus $\alpha =1$.
As for $n_i$, if  $\Phi_i$
originates from $D7$-branes, we have $n_i=0$ \cite{ibanez1}.
On the other hand, $n_i=1/2$ for the matter fields living on
the intersections of $D7$ branes, and $n_i=1$ for the matter fields
living on either the triple intersection of $D7$ branes or $D3$ branes
\cite{ibanez1}.
Our approach here is not to consider a specific string compactification,
but to consider generic effective SUGRA models
described by arbitrary rational numbers
$n_0, n_P$ and $n_i$ which are being of order unity.

Mixed modulus-anomaly mediation can give a low energy sparticle
spectrum which is quite different from other scenarios of SUSY breaking.
Taking into account 1-loop RG evolution,
the low energy gaugino masses are given by
\bea
\label{lowgaugino}
M_a(\mu)&=&M_0\left[
1-\frac{1}{4\pi^2}b_ag_a^2(\mu)\ln\left(\frac{M_{GUT}}{(M_{Pl}/m_{3/2})^{\alpha/2}\mu}\right)
\right],
\eea
where $g_a(\mu)$ are the running gauge couplings at scale $\mu$.
The low energy values of $A_{ijk}$ and $m_i^2$ generically
depend on the associated Yukawa couplings $y_{ijk}$.
However if
$y_{ijk}\lesssim 1/\sqrt{8\pi^2}$,
{\it or}  the following  conditions are satisfied for
the $i$-$j$-$k$  combination with $y_{ijk}\gtrsim 1/\sqrt{8\pi^2}$:
\bea
\label{con2}
\frac{\tilde{A}_{ijk}}{M_0}
=\frac{\tilde{m}_i^2+\tilde{m}_j^2+\tilde{m}_k^2}{M_0^2}
= 1,
\eea
their low energy values are given by
\bea
\label{lowsoft}
&& A_{ijk}(\mu)\,= \,\tilde{A}_{ijk}
+\frac{M_0}{8\pi^2}(\gamma_i(\mu)
+\gamma_j(\mu)+\gamma_k(\mu))
\ln\left(\frac{M_{GUT}}{(M_{Pl}/m_{3/2})^{\alpha/2}\mu}\right),
\nonumber \\
&& m_i^2(\mu)\,= \,
\tilde{m}_i^2
-\frac{1}{8\pi^2}Y_i\Big(\sum_j\tilde{m}^2_jY_j\Big)
g_Y^2(\mu)\ln\left(\frac{M_{GUT}}{\mu}\right)
\nonumber \\
&&\,\,\, +\,\,\frac{M_0^2}{4\pi^2}
\left\{\gamma_i(\mu)-\frac{1}{2}\frac{d\gamma_i(\mu)}{d\ln\mu}
\ln\left(\frac{M_{GUT}}{(M_{Pl}/m_{3/2})^{\alpha/2}\mu}\right)\right\}
\ln\left(\frac{M_{GUT}}{(M_{Pl}/m_{3/2})^{\alpha/2}\mu}\right)
,
\eea
where $\gamma_i(\mu)$
denote the running anomalous dimensions
at $\mu$ and $Y_i$ is the $U(1)_Y$ hypercharge
of $\Phi_i$.



The results of (\ref{lowgaugino}) and (\ref{lowsoft})  show an interesting feature:
when $\sum_i \tilde{m}_i^2Y_i=0$ and the condition (\ref{con2})
is satisfied,
the low energy soft masses in the
mixed modulus-anomaly mediation with messenger scale
$M_{GUT}$ are same as
the low energy soft masses in the pure modulus-mediation
started from the {\it mirage messenger scale}
$\Lambda_{m} \approx (m_{3/2}/M_{Pl})^{\alpha/2}M_{GUT}$.
This feature can have an interesting
implication for the little hierarchy between the Higgs  masses
and the other superparticle masses in the MSSM.
To see this, let us consider a class of models with
$n_0, n_i$ and $n_P$ satisfying
\bea
\label{condition_n}
n_P&=& \frac{n_0}{3},\quad
n_{H_u}\,=\,n_{H_d}\,=\,\frac{n_0}{3},
\nonumber \\
\sum_i n_iY_i&=& 0,
\quad
n_0-n_{H_u}-n_{Q_3}-n_{U_3}\,=\, 1,
\eea
for which
\bea
\label{condition1}
&&\alpha=2,
\quad \tilde{m}_{H_u}^2=\tilde{m}_{H_d}^2=0,
\quad
\sum_i\tilde{m}^2_iY_i=0,
\nonumber \\
&&\frac{\tilde{A}_t}{M_0}=\frac{\tilde{m}_{H_u}^2+\tilde{m}_{Q_3}^2
+\tilde{m}^2_{U_3}}{M_0^2}=1,
\eea
where $A_t=A_{H_uQ_3U_3}$ for  $Q_3$ and $U_3$
denoting the top-doublet and
the top-singlet.
Note that under the assumption that $n_0,n_i$ and $n_P$ are rational
numbers, the conditions of (\ref{condition_n}) are not a parameter fine-tuning,
but correspond to a restriction to the specific class of models.
For these models, the low energy
expressions (\ref{lowgaugino}) and (\ref{lowsoft})
are  applicable.
(As we will see, such models have
a low $\tan\beta\lesssim 5$, for which
the $b$ and $\tau$ Yukawa couplings  $y_{b,\tau}\lesssim
1/\sqrt{8\pi^2}$.)
Then one easily finds that
the model predicts
\bea
\label{prediction}
M_a(\Lambda_{m})&=& M_0\Big[1+{\cal O}(1/8\pi^2)\Big],
\quad
A_t(\Lambda_{m})\,=\, M_0\Big[1+{\cal O}(1/8\pi^2)\Big],
\nonumber \\
m_{\tilde{t}_L}^2(\Lambda_{m})+m_{\tilde{t}_R}^2(\Lambda_m)&=&M_0^2
\Big[1+{\cal O}(1/8\pi^2)\Big],
\quad
m_{H_{u,d}}^2(\Lambda_{m})\,=\, {\cal O}(M_0^2/8\pi^2),
\eea
where $\Lambda_{m}\sim M_{GUT}m_{3/2}/M_{Pl}\sim 1$ TeV,
thus realizes the Higgs-stop little hierarchy
$m_H^2/m_{\tilde{t}}^2={\cal O}(1/8\pi^2)$ in a natural manner.

Unfortunately, the precise low energy values of $m_{H_{u,d}}^2$
are sensitive to the unknown threshold corrections of
${\cal O}(M_0^2/8\pi^2)$
at $M_{GUT}$ (or at $M_{st}$) as well as the higher loop
RG effects below $M_{GUT}$ and the SUSY threshold corrections at TeV scale.
It is still conceivable that $m_{H_u}^2$ is negative
at the weak scale.
For instance, $m_{H_u}^2$ at the weak scale
might be dominated by the radiative corrections below $\Lambda_m$,
$\Delta m_{H_u}^2 \sim -\frac{3y_t^2}{4\pi^2}
m_{\tilde t}^2 \ln ({\Lambda_{m}}/{m_{\tilde t}})$,
if $\Lambda_{m}$ is somewhat bigger than $m_{\tilde t}$.
Then $m_{H_u}^2/m_{\tilde{t}}^2$ at the weak scale
would be negative and bigger than $1/8\pi^2$ by a factor of few.
Although not essential, one might choose also $n_{Q_3}=n_{U_3}$
for which $m_{\tilde{t}_L}^2(\Lambda_m)\simeq m_{\tilde{t}_R}^2(\Lambda_m)
\simeq M_0^2/2$ and thus the radiative correction to $m_{h^0}$ becomes maximal.

In the above model, the little hierarchy $m_H^2/m_{\tilde{t}}^2={\cal O}(1/8\pi^2)$
could be obtained  since
the RG evolution of $m_H^2$ between $M_{GUT}$ and $\Lambda_m\sim 1$ TeV
is canceled by the anomaly-mediated contribution at $M_{GUT}$.
All conditions in (\ref{condition1})
are necessary for the little hierarchy
$m_{H_{u,d}}^2={\cal O}(m_{\tilde{t}}^2/8\pi^2)$ to be achieved
through such cancellation.
We stress that the cancellation of RG evolution and the associated low energy predictions
(\ref{prediction}) are
the inevitable consequences of any SUSY breaking scenario
yielding the soft terms of the form (\ref{soft1}) satisfying
(\ref{condition1}), which could be naturally realized
in KKLT-motivated effective SUGRA.
Note that while the gaugino masses appear to be unified
at TeV, the corresponding gauge couplings  are still
unified at $M_{GUT}\sim
2\times 10^{16}$ GeV.

In order for the model to be viable, one needs also that
the other two Higgs mass parameters $\mu$ and $B$
satisfy the conditions for electroweak symmetry breaking:
\bea
\label{ewbreaking}
|B\mu|^2 &>& (m_{H_u}^2+|\mu|^2)(m_{H_d}^2+|\mu|^2),
\nonumber \\
2|B\mu| &<&m_{H_u}^2+m_{H_d}^2+2|\mu|^2,
\eea
for the Higgs potential
\bea
V_{\rm higgs}&=&
(m_{H_u}^2+|\mu|^2)|H_u^0|^2+
(m_{H_d}^2+|\mu|^2)|H_d^0|^2
\nonumber \\
&&-\,
(B\mu H^0_uH^0_d+{\rm c.c.})
+\frac{1}{8}(g_1^2+g_2^2)
(|H_u^0|^2-|H_d^0|^2)^2.
\eea
The parameter $|\mu^2|$ must be of ${\cal O}(m^2_{H_u})={\cal O}(M_Z^2)$ to
avoid the fine-tuning in Eq.~(\ref{MZ-mu-mHud}).
Then the above electroweak symmetry breaking conditions would require
\bea
\mu \,\sim \, B \,\sim\, m_{H_{u,d}} \,\sim\, M_0/\sqrt{8\pi^2},
\eea
which appears to be difficult to be realized  in the mixed modulus-anomaly
mediation.
In fact,
$B$ in the mixed modulus-anomaly mediation is generically of the order of $m_{3/2}
\sim 4\pi^2M_0$,
which is obviously too large to allow the electroweak symmetry breaking.
However for the models with $n_P=n_0/3$ yielding $\alpha=2$,
one can achieve the desired size of $\mu$ and $B$ without
severe fine-tuning.

To see this, let us assume that the Higgsino mass parameter
$\mu$ is generated by the same non-perturbative dynamics
stabilizing $T$, thus the superpotential contains \cite{chun}
\bea
\Delta W \,=\, \tilde{A}e^{-aT}H_uH_d.
\eea
The resulting $\mu$ and $B$ for the canonically normalized Higgs doublets
renormalized at scales just below $M_{GUT}$ are given by
\bea
\label{b_gut}
\mu &=& \frac{e^{K_0/2}\tilde{A}e^{-aT}}{\sqrt{Z_{H_u}Z_{H_d}}},
\nonumber \\
B&=&M_0\left[a(T+T^*)+(T+T^*)\partial_T\ln (e^{-2K_0/3}Z_{H_u}Z_{H_d})\right]
\nonumber \\
&& -\frac{F^C}{C_0}+\frac{1}{16\pi^2}(\gamma_{H_u}+\gamma_{H_d})\frac{F^C}{C_0},
\eea
where
$\gamma_{H_{u,d}}$ denote the anomalous dimension of the Higgs doublets.
One can simply choose the free parameter $\tilde{A}$ to take a value yielding
$\mu(\Lambda_m)\sim m_H$, which
does not interfere with
the other parts of the model and
is technically natural.
On the other hand, $B$
contains $F^C/C_0$ and $a(T+T^*)M_0$  which are of the order of $m_{3/2}
\approx 4\pi^2 M_0$,
thus too large in general.
(Note that $aT\sim 4\pi^2$ in Eq.~(\ref{vev}).)
However for the models with
$n_P=n_0/3$, these two contributions
of ${\cal O}(m_{3/2})$
cancel to each other, leaving only a piece of ${\cal O}(M_0)$.
In fact, the precise value of $B$ is sensitive to the
unknown higher order stringy or loop-threshold corrections to
$K_0$ and ${\cal P}_{\rm lift}$ which can be parameterized as
\bea
K_0&=&-n_0\ln(T+T^*) +\Delta K_0,
\nonumber \\
\ln ({\cal P}_{\rm lift}/D) &=&n_P\ln (T+T^*) +\Delta \Omega.
\eea
Including the effects of $\Delta K_0$ and $\Delta\Omega$,
we find that the low energy value of $B$ in the models of
(\ref{condition_n}) is given by
\bea
\label{b_parameter}
B(\Lambda_m)&=&a(T+T^*)M_0-\frac{F^C}{C_0} +{\cal O}(M_0/8\pi^2)
\nonumber \\
&=& M_0\left[
a(T+T^*)\left(1-\frac{\partial_TK_0}{3\partial_T\ln(V_{\rm lift})}\right)
-\frac{2n_0}{3}+{\cal O}\left(\frac{1}{8\pi^2}\right) \right]
\nonumber \\
&=& -M_0\left[
\frac{a(T+T^*)}{n_0}(T+T^*)\partial_T\Big(
3\Delta\Omega-\Delta K_0\Big)
+\frac{2n_0}{3}+{\cal O}\left(\frac{M_0}{8\pi^2}\right)\right],
\eea
for $n_P=n_0/3$ yielding $\alpha=2$. Note that
the last term of $B(M_{GUT})$ in (\ref{b_gut}) cancels the RG evolution
of $B$ down to $\Lambda_m\sim M_{GUT}m_{3/2}/M_{Pl}\sim 1$ TeV,
thereby $B(\Lambda_m)$ is simply determined by $a(T+T^*)M_0$ and $F^C/C_0$
as in (\ref{b_parameter}).
Since $a(T+T^*)\sim 8\pi^2$, the part depending on
$\Delta K_0$ and $\Delta \Omega$
can be important even when  $\Delta K_0$ and/or $\Delta \Omega$
are the corrections of ${\cal O}(1/8\pi^2)$.
With a minor tuning of such higher order effects,
e.g. a tuning of $10\sim 20$ \%, one can obtain
$B(\Lambda_m)$ which is small enough, e.g. $0.2M_0$, to satisfy the
electroweak symmetry breaking condition
(\ref{ewbreaking})
for $\mu\sim m_H\sim M_0/\sqrt{8\pi^2}$.

As summarized in (\ref{prediction}),
the models of (\ref{condition_n}) give  strong predictions
on the sparticle masses.
The model predicts
approximately universal low energy gaugino masses,
$M_a(\mbox{TeV})\simeq M_0 ={\cal O}(1)$ TeV, which is  the consequence of
$n_P=n_0/3$ yielding $\alpha=2$,
and also the stop masses satisfying
$m^2_{\tilde{t}_L}(\mbox{TeV})+m^2_{\tilde{t}_R}(\mbox{TeV})\simeq M_0^2$.
In view of Eq.~(\ref{b_parameter}), it is difficult that $B$ is significantly smaller than
$\mu\sim m_{H_{u,d}}$, thus the resulting  $\tan\beta$
is expected to be less than moderate, e.g. $\tan\beta\lesssim 5$, justifying
our assumption $y_{b,\tau}\lesssim 1/\sqrt{8\pi^2}$.
Obviously then the LSP of the model is almost
Higgsino-like neutralino, and the next LSP is the almost
Higgsino-like chargino.

It is quite remarkable that the model discussed above
naturally avoids the SUSY flavor and CP problems
as well as giving the little hierarchy $m_{H}^2\sim m_{SUSY}^2/8\pi^2$.
First of all, the model is free from dangerous SUSY flavor
violation if $n_i$ are chosen to be flavor-independent,
which is a rather plausible possibility in view of their stringy
origin.
As for SUSY CP, it has been noticed \cite{choi1} that
the non-linear PQ symmetry (\ref{u1t})
of the model guarantees that one can always choose a field basis
in which $M_0$ and $m_{3/2}$ are real,
and thus $M_a$ and $A_{ijk}$ are real also.
The result of (\ref{b_parameter}) shows that
the invariance of $\Delta K$ and $\Delta\Omega$ under
$U(1)_T$ assures that $B$ is real also in the same
field basis, thus the model is completely free from
dangerous SUSY CP violation.

There is another interesting aspect
of the model related to the color and/or charge breaking (CCB) and
the unbounded-from-below (UFB) direction in the full scalar potential.
Detailed studies on CCB and UFB directions of the MSSM  potential
have been carried out in Ref.~\cite{Casas:1995pd}.
The most serious constraint is obtained by the
so-called UFB-3 direction, which includes the up-sector Higgs and
slepton fields $\{H_u,\tilde \nu_{L_i},\tilde e_{L_j},
\tilde e_{R_j}\}$.
The potential along the UFB-3 direction becomes
unbounded-from-below if
$m_{H_u}^2 + m_{\tilde L_i}^2 <0$ at low energy scale.
In many models,
$m_{H_u}^2 + m_{\tilde L_i}^2$ at low-energy scale becomes
negative
because  $\Delta m^2_{H_u} \sim - m^2_{\tilde t}$ and
$m^2_{\tilde t}\gg m_{\tilde L_i}^2$
due to the RG evolution effects from  gluino mass.
However, for the class of models discussed above,
one can easily arrange $n_i$ to get
$m_{\tilde L_i}^2(\Lambda_{m}) \sim M_0^2$ and thus
 $m_{H_u}^2 + m_{\tilde L_i}^2> 0$ at low energy scale.

So far, we have been discussing  the models leading to
$m^2_{H_u}(\Lambda_{m}) \sim m^2_{H_d}(\Lambda_{m})\sim
m_{SUSY}^2/8\pi^2$.
One might consider an alternative scenario leading to
the different pattern of little hierarchy:
\bea
\label{model2}
m^2_{H_d}(\Lambda_{m}) \sim  8\pi^2 m^2_{H_u}(\Lambda_{m})\sim m_{SUSY}^2.
\eea
In order to get such pattern of low energy spectrum, one
still needs the last condition of (\ref{condition_n})
as well as $n_P=n_{H_u}=n_0/3$.
Concerned about $n_{H_d}$, we need more conditions:
\bea
\label{con_model2}
n_0-1 &=& n_{H_d}+n_{Q_3}+n_{D_3}
=n_{H_d}+n_{L_3}+n_{E_3},
\nonumber \\
n_{H_d} &<& \frac{n_0}{3},
\qquad
\frac{1}{2}\left(n_{H_d}-\frac{1}{3}n_0\right) \,=\,\sum_{\rm matter} n_iY_i,
\eea
where the first condition
is to satisfy (\ref{con2}) for the
$b$ and $\tau$ Yukawa couplings which are not negligible anymore
since $\tan\beta \sim \sqrt{8\pi^2}$ (see the discussion below),
the second condition is required for
$m^2_{H_d}(\Lambda_{m}) = {\cal O}(M_0^2)$,
and the last condition is introduced to protect small $m_{H_u}^2$ from
the RG running effect proportional to $\sum_i \tilde{m}_i^2Y_i$.
In such models, $B = {\cal O}(M_0)$ would
satisfy the symmetry breaking condition
(\ref{ewbreaking}), and
the expected  ratios of the Higgs mass parameters at the weak scale are given by
\begin{equation}
m^2_{H_d} : |m^2_{H_u}| : |\mu B| : \mu^2 =
{\cal O}(M^2_0) : {\cal O}(M^2_0/{8 \pi^2}) :
 {\cal O}(M^2_0/\sqrt{8 \pi^2}):
{\cal O}(M^2_0/{8 \pi^2}) .
\end{equation}
The resulting $\tan \beta$ is determined by
\begin{equation}
\frac{ \tan \beta}{1+ \tan^2 \beta} =
\frac{\mu B}{m_{H_d}^2 + m_{H_u}^2+2 \mu^2} =
{\cal O}\left(\frac{1}{\sqrt{8\pi^2}}\right),
\end{equation}
yielding a moderately large $\tan \beta
={\cal O}(\sqrt{8\pi^2})$.
The above models for the mass pattern (\ref{model2}) appear to be
less attractive than
the models of  (\ref{condition_n}) yielding $m_{H_{u}}^2\sim
m_{H_d}^2 \sim M_0^2/8\pi^2$ as they require more conditions
on $n_i$.



To summarize, we pointed out that
the little hierarchy
$m_H^2\sim m_{SUSY}^2/8\pi^2$
which is desirable for the lightest MSSM higgs boson
to satisfy the experimental bound  can be
naturally realized in 4D effective  SUGRA models with certain class of
uplifting potential.
Such spectrum is realized by the cancellation
between the anomaly mediated soft terms at $M_{GUT}$
and the subsequent RG evolution down to the TeV scale.
Under a reasonable condition, the model
can give rise to  $\mu$ and $B$ satisfying the
electroweak symmetry breaking condition
without severe fine-tuning.
Furthermore, the model naturally avoids
dangerous  SUSY flavor and
CP violations, and also is favorable from the
viewpoint of CCB and UFB constraints.

The model
predicts a unique  low-energy spectrum.
The three MSSM gaugino masses $M_a$ ($a=1,2,3$)
are almost universal at low energy scale, $M_a(\mbox{TeV}) \simeq M_0
={\cal O}(1)$ TeV,
and the stop masses satisfy the sum rule
$m^2_{\tilde{t}_L}(\mbox{TeV})+m^2_{\tilde{t}_R}(\mbox{TeV})\simeq M_0^2$.
The LSP is the Higgsino-like neutralino with a mass of ${\cal O}(100)$ GeV,
the next LSP is the Higgsino-like chargino, and
their masses are nearly degenerate.
The gravitino mass $m_{3/2}\sim 4\pi^2 M_0 \sim {\cal O}(30)$ TeV
and the modulus mass $m_T \sim 8\pi^2 m_{3/2}={\cal O}(10^3)$ TeV,
thus can avoid the cosmological gravitino/moduli problem \cite{Endo:2005uy}.
It would be quite interesting to study
more phenomenological aspects of our model as
well as the cosmological aspects \cite{cjko}.

\vspace{5mm}
\noindent{\large\bf Acknowledgments}
\vspace{5mm}

K.C. and K.S.J are supported by the KRF Grant funded by
the Korean Government (KRF-2005-201-C00006),
the KOSEF Grant (KOSEF R01-2005-000-10404-0),
and the Center for High Energy Physics of Kyungpook National University.
T.K.\/ is supported in part by the Grand-in-Aid for Scientific
Research \#16028211 and \#17540251, and
the Grant-in-Aid for
the 21st Century COE ``The Center for Diversity and
Universality in Physics''
and
K.O. is supported by the Grand-in-aid for Scientific Research on
Priority Areas \#441: ``Progress in elementary particle physics of the
21st century through discovery of Higgs boson and supersymmetry''
 \#16081209
from the Ministry of Education, Culture,
Sports, Science and Technology of Japan.

\end{document}